\documentstyle[eqsecnum,floats,preprint,epsf,aps]{revtex}
\tighten

\begin{document}

\renewcommand{\arraystretch}{1.5}
\newcommand{\be}{\begin{equation}}
\newcommand{\ee}{\end{equation}}
\newcommand{\bea}{\begin{eqnarray}}
\newcommand{\eea}{\end{eqnarray}}
\def\Tr{\mathop{\rm Tr}\nolimits}
\def\mapright#1{\smash{\mathop{\longrightarrow}\limits^{#1}}}
\def\mapdown#1{\big\downarrow \rlap{$\vcenter
  {\hbox{$\scriptstyle#1$}}$}}
\def\su#1{{\rm SU}(#1)}
\def\so#1{{\rm SO}(#1)}
\def\sp#1{{\rm Sp}(#1)}
\def\u#1{{\rm U}(#1)}
\def\o#1{{\rm O}(#1)}
\def\p#1{{\pi_{#1}}} 
\def\Z{{\bf Z}}
\def\R{{\bf R}}
\def\M{{\cal M}}
\def\L{{\cal L}}
\def\c{c_\chi}
\def\s{s_\chi}
\def\xh{\hat x}
\def\yh{\hat y}
\def\zh{\hat z}
\def\vt{\widetilde{v}}
\def\wt{\widetilde{w}}
\def\n#1{{\hat{n}_{#1}}}
\def\c#1{\cos{#1}}
\def\s#1{\sin{#1}}
\def\cs#1{\cos^2{#1}}
\def\ss#1{\sin^2{#1}}
\newcommand{\PSbox}[3]{\mbox{\rule{0in}{#3}\includegraphics{#1}\hspace{#2}}}

\title{Dirichlet Topological Defects}

\author{Sean M. Carroll$^{1}$\footnote[1]{\tt carroll@itp.ucsb.edu} and 
Mark Trodden$^{2}$\footnote[2]{\tt trodden@theory1.physics.cwru.edu.}}

\address{~\\$^1$Institute for Theoretical Physics \\
University of California \\
Santa Barbara, California 93106, USA.}

\address{~\\$^2$Particle Astrophysics Theory Group\\ 
Department of Physics \\
Case Western Reserve University \\
10900 Euclid Avenue \\
Cleveland, OH 44106-7079, USA.}

\maketitle

\begin{abstract}
We propose a class of field theories featuring solitonic solutions
in which topological defects can end when they intersect other defects of
equal or higher dimensionality.  Such configurations may be termed
``Dirichlet topological defects'', in analogy with the D-branes of
string theory.  Our discussion focuses on defects in scalar field
theories with either gauge or global symmetries, in (3+1) dimensions;
the types of defects considered include walls ending on walls, strings
on walls, and strings on strings.
\end{abstract}

\setcounter{page}{0}
\thispagestyle{empty}

\vfill

\noindent NSF-ITP/97-117\hfill hep-th/9711099

\noindent CWRU-P16-97


\eject

\vfill

\eject

\baselineskip 20pt plus 2pt minus 2pt

\section{Introduction}
\label{sec:introduction}

It has long been appreciated that field theories with
spontaneously broken symmetries often support topological defects 
--- solitonic solutions whose stability is guaranteed by a topological
conservation law.  A variety of such defects can occur, depending
on the pattern of symmetry breaking in the model.  When a symmetry
group $G$ is spontaneously broken to a subgroup $H$, the types of
defects supported depend on the homotopy properties
of the vacuum manifold, $\M = G/H$.
In a $(d+1)$-dimensional spacetime, $p$-dimensional defects ($p<d$)
exist if the homotopy group $\p{d-p-1}(\M)$ is nontrivial.  Thus,
in 3 spatial dimensions there will be planar defects (domain
walls) if $\p0(\M)\neq 0$, line-like defects (cosmic strings)
if $\p1(\M)\neq 0$, and pointlike defects (monopoles) if 
$\p2(\M)\neq 0$.  (For reviews see \cite{reviews}.)

In addition to these basic defects, there are various
composite solutions which combine two of the types, generally when
a $(p-1)$-dimensional defect serves as the boundary of a $p$-dimensional
defect.  Consider, for example, a sequence of symmetry breakings in
which a group $G$ is broken to $H\times\Z_2$ at some high scale, and 
subsequently to $H$ at some lower scale, where both $G$ and $H$ 
are connected and simply connected.  Then $\p1[G/(H\times\Z_2)]=\Z_2$,
and strings will be formed at the first phase transition.  A 
closed loop in physical space around the string, parameterized 
by $x^\mu(\lambda)$ as $\lambda$ goes from $0$ to $1$, defines a
closed loop in field space which can be written $\phi(\lambda) = 
g(\lambda)\phi_0$, where $\phi_0$ is the initial value of the field
and $g(\lambda)$ is a path in the group $G$ such that
$g(0) = 1$.  The group element
$g(1)$ corresponds not to the identity in $G$, but to the
non-identity element in the unbroken $\Z_2$ subgroup.  When this
group is broken after the second phase transition, the path
$\phi(\lambda)$ no longer describes a closed loop in the vacuum
manifold; rather, a domain wall must form with the string as its
boundary \cite{kls}.  Looked at another way, if we were unaware of the full 
symmetry group $G$, we would predict the appearance of topologically
stable walls due to the breaking $H\times\Z_2\rightarrow H$; but
in fact the presence of $G$ implies that these walls can end on cosmic 
strings.  Indeed, such walls
are unstable to the nucleation of holes bounded by string loops (although
the timescale for such processes may be extraordinarily long). 
Similarly, models can be constructed \cite{stringmonopole}
in which strings end on monopoles,
and can decay via the nucleation of monopole-antimonopole pairs along 
their length.  Finally, in some theories the evolution of one kind of
defect can lead to the creation or destruction of another kind;
examples include domain walls sweeping up monopoles \cite{dlv}
and collapsing textures nucleating monopoles or string loops
\cite{textures}.

In this paper we discuss configurations which are complementary to
those mentioned above --- solutions in which defects can end when they
intersect other defects of equal or higher dimensionality, such as
strings ending on domain walls.  For such configurations, we term the
defects on which other defects end ``Dirichlet topological
defects'', in analogy with the D-branes of string theory.  The latter
are extended objects on which fundamental strings can end
\cite{horava,db1,jp,db2}.  The models considered here are ordinary 
field theories, which support topological solitons which resemble these
objects in fundamental string theory.  (Dirichlet defects in this
sense have been discussed previously:  cosmic strings ending on domain
walls can arise in supersymmetric Yang-Mills theories \cite{qcd} as well 
as in grand unified models \cite{dlv}, while the well-known phenomenon
of non-intercommuting cosmic strings provides an example of strings
ending on strings.)  We will focus on the case of defects
in bosonic field theories in (3+1) dimensions; generalization to
higher dimensions and theories with fermions is left to future work.

\section{Walls Ending on Walls}
\label{wallsandwalls}

In any number of spatial dimensions, the simplest example of Dirichlet 
topological defects consists of codimension-one defects ending
on other codimension-one defects: for $d=3$, domain walls ending on
domain walls.  This example provides a paradigm for the
models considered in subsequent sections.

Walls arise upon the breakdown of discrete symmetries, and it is
therefore unnecessary to introduce gauge fields into our model at
this stage.  To form the Dirichlet wall (or D-wall) we introduce 
a single real scalar $\phi$, invariant under a symmetry group
$\Z_2^{(1)}$ which acts on $\phi$ via $\phi \rightarrow -\phi$.  
If the potential energy is minimized at $\phi = \pm v$, 
the wall solution interpolates from one domain with $\langle\phi\rangle 
= v$ to another with $\langle\phi\rangle=-v$.  
We next introduce another real scalar $\psi_1$, invariant under a
distinct symmetry group $\Z_2^{(2)}$ which sends $\psi_1$ to $-\psi_1$.
In order for $\psi_1$ to lead to ``fundamental'' walls which can end
on the D-wall, we require that the potential be minimized at
$\psi_1 = \pm w$ when $\phi = v$, and $\psi_1 = 0$ when $\phi = -v$.
Such a potential is not invariant under the original symmetry
$\Z_2^{(1)}$ unless we include a third real scalar $\psi_2$ which
exchanges roles with $\psi_1$ under the action of $\Z_2^{(1)}$.  That is,
we consider a complete symmetry group 
$\Z_2^{(1)}\times \Z_2^{(2)}\times \Z_2^{(3)}$, 
with action
\begin{eqnarray}
 & \Z_2^{(1)}: & \ \ \{\phi \rightarrow -\phi\ ,\ \psi_1 \leftrightarrow 
 \psi_2\} \ , 
 \nonumber\\
 & \Z_2^{(2)}: & \ \ \psi_1  \rightarrow -\psi_1\ , 
 \label{symm1}\\
 & \Z_2^{(3)}: & \ \ \psi_2  \rightarrow -\psi_2\ . 
  \nonumber
\end{eqnarray}
Such a symmetry allows for $\psi_1$-walls when $\langle\phi\rangle 
= v$ and $\psi_2$-walls when $\langle\phi\rangle=-v$; each can end on
the Dirichlet walls where $\phi$ changes values.  

Given the three scalar fields $\{\phi, \psi_1, \psi_2\}$, a complete
set of nonderivative interactions of no higher than fourth order
which are consistent with these symmetries includes $\phi^2$, $\phi^4$, 
$(\psi_1^2 + \psi_2^2)$, $(\psi_1^4 + \psi_2^4)$, $\psi_1^2\psi_2^2$,
$\phi^2(\psi_1^2 + \psi_2^2)$, and $\phi(\psi_1^2 - \psi_2^2)$.  It is
convenient to write the most general potential constructed from these
terms in the form 
\begin{eqnarray} 
  V(\phi,\psi_1,\psi_2) 
  & = & \lambda_\phi(\phi^2 - \vt^2)^2 + \lambda_\psi\left[\psi_1^2
  + \psi_2^2 - \wt^2 + g(\phi^2  - \vt^2)\right]^2
  \nonumber\\
  & & \quad {}  + h \psi_1^2 \psi_2^2- \mu\phi(\psi_1^2 - \psi_2^2) \ .
  \label{pot}
\end{eqnarray}
We consider the parameter space in which $\lambda_\phi$, $\lambda_\psi$,
$h$, $\vt^2$ and $\wt^2$ are all positive.  The sign of $g$ is
left unspecified, and we may take $\mu\geq 0$ without loss of generality
(as a change in sign of $\mu$ is equivalent to a field redefinition
interchanging $\psi_1$ and $\psi_2$).
At $\mu=0$, the potential is the sum of three non-negative terms
which may be simultaneously set to zero.  It is therefore
easy to see that there exist eight vacua, in which
$\phi = \pm \vt$ and $(\psi_1,\psi_2)$ are either $(\pm\wt, 0)$ or
$(0, \pm\wt)$.  
As $\mu$ is increased to positive values, this degeneracy
is broken, and it becomes energetically favorable (for example)
to have $|\psi_1| = w = \wt +{\cal O}(\mu)$ and $|\psi_2| = 0$ when
$\phi = v = \vt +{\cal O}(\mu)$.

\begin{figure}
\vskip -9cm
\centerline{\epsfbox{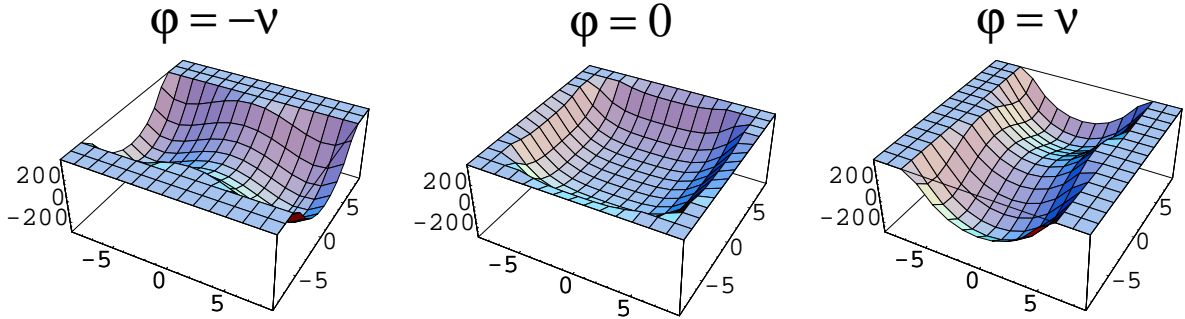}}
\vskip -9cm
\vspace{0.3in}
\caption{\sf Shape of the potential as a function of
  $\psi_1$ (right horizontal axis) and $\psi_2$ (left horizontal axis), 
  for three different values of $\phi$.  When $\phi = v$,
  $\langle\psi_1\rangle = \pm w$ and $\langle\psi_2\rangle = 0$, while 
  when $\phi = -v$,
  $\langle\psi_1\rangle = 0$ and $\langle\psi_2\rangle = \pm w$.}
\label{wallplot}
\end{figure}

The potential in the vicinity of its minima is plotted in
Fig.~\ref{wallplot}.
There are four vacua of zero energy, of the form 
$\{\phi = v,\psi_1 = \pm w,\psi_2 = 0\}$ and
$\{\phi = -v,\psi_1 = 0,\psi_2 = \pm w\}$.  
The vacuum expectation value (vev) $\langle |\phi |\rangle=v$ solves a
cubic equation
\begin{equation} 
  8\lambda_\phi\lambda_\psi v^3 - 6\lambda_\psi g\mu v^2 
  - (8\lambda_\phi\lambda_\psi\vt^2 + \mu^2)v 
  - 2\lambda(g\vt^2 + \wt^2)\mu  =0\ .
\end{equation} 
The correct root of this equation is the one which reduces to
$\vt$ at $\mu = 0$.
The vev $w$ is given in terms of $v$ by
\begin{equation} 
  w = \left(\wt^2 + g(\vt^2 - v^2) -
  {\mu v \over{2\lambda_\psi}}\right)^{1/2}
  \ .
  \label{vev2}
\end{equation} 
In any of these vacua, the original $\Z_2\times\Z_2\times\Z_2$ is
broken to a single $\Z_2$ and the vacuum manifold is therefore $\M = 
(\Z_2\times\Z_2\times\Z_2)/\Z_2 = \Z_2\times\Z_2$.  Starting from
a single vacuum we can visit three different
vacua, two of which are associated with Dirichlet walls of equal energy,
and the third of which is associated with a fundamental wall.

The energies of the two types of wall are complicated
functions of the parameters in the potential.  However, there is
a simple argument that provides an upper limit on the energy density 
of the fundamental wall in comparison to the D-wall:
\begin{equation} 
  E_F \leq 2 E_D\ .
\end{equation}
To see this, consider the energy of the fundamental
wall as a functional of the wall
profile. This is determined by the values of $\phi(x)$, $\psi_1(x)$ and 
$\psi_2(x)$ as
$x$ takes values from $x_1$, where the fields are in (for example) the
vacuum $\{ \phi = v, \psi_1 = w, \psi_2 = 0\}$, to $x_2$ where
the fields are in the vacuum $\{ \phi = v, \psi_1 = -w, \psi_2 = 0\}$.
This energy (per unit area) is given by
\begin{equation} 
  E_F = \int_{x_1}^{x_2} \left[{1\over 2}(\nabla\phi)^2 + 
  {1\over 2}(\nabla\psi_1)^2 + {1\over 2}(\nabla\psi_2)^2 +
  V(\phi,\psi_1,\psi_2)\right]\, dx\ . 
\end{equation}
The stable wall configuration is that which minimizes this
energy.  However, there are paths in field space which go through the
intermediate point $\{ \phi = -v, \psi_1 = 0, \psi_2 = w\}$ and thus
correspond to configurations representing two parallel
D-walls.  There may be (and typically will be) configurations
with lower energy than this one, so the energy of the fundamental
walls is bounded above by the energy of two D-walls.

It would appear to be very difficult to find exact solutions
representing a fundamental wall ending on a D-wall.  Not only 
do the configurations of the isolated walls involve the interactions
of all three fields, but the D-wall is not expected to be smooth
at the point where it is intersected by a fundamental wall; the tension
from the latter will pull the D-wall into a cusp configuration.
However, it may be possible to find solutions to an effective world-volume
theory of the walls, and work in this direction has been undertaken for
the case of D-branes in fundamental string theory \cite{pulling}.

\section{Strings Ending on Walls}
\label{stringsandwalls}

The case of strings ending on D-walls in three
spatial dimensions is an immediate generalization of the previous
example.  Strings arise most simply from the breakdown of $\u1$
symmetries; we therefore promote the real scalars $\psi_1$ and $\psi_2$
to complex fields $\psi_i = \rho_i e^{i\theta_i}$,
and the symmetries acting on them to $\u1$'s,
leaving the discrete $\Z_2$ (which breaks to give the D-walls)
unchanged.  The complete set of symmetries is therefore

\begin{eqnarray}
 \Z_2:&\ \ \ &\{\phi \rightarrow -\phi\ ,\ \psi_1 \leftrightarrow \psi_2\} \ ,
 \nonumber\\
 {\u1}_1:&\ \ \ & \psi_1  \rightarrow e^{-i\omega_1}\psi_1\ , 
 \label{symm2}\\
 {\u1}_2:&\ \ \ & \psi_2  \rightarrow e^{-i\omega_2}\psi_2\ .
  \nonumber
\end{eqnarray}
The two $\u1$'s may be taken to be either global or gauge symmetries. 
In the latter case,
$\omega_1$ and $\omega_2$ are functions of spacetime, and
there are two gauge fields $A_\mu^{(1)}$, $A_\mu^{(2)}$, with
the usual transformation properties
\begin{equation} 
  A_\mu^{(i)} \rightarrow A_\mu^{(i)} + \partial_\mu \omega_i
\end{equation}
and associated covariant derivatives
\begin{equation} 
  D_\mu\psi_i = \partial_\mu\psi_i +i A_\mu^{(i)}\psi_i\ .
  \label{covderiv}
\end{equation}
Since the real scalar $\phi$ is uncharged under both $\u1$'s,
the kinetic term for $\phi$ is given in terms of
ordinary partial derivatives.

The appropriate potential is now
\begin{eqnarray} 
  V(\phi,\psi_1,\psi_2) 
  & = & \lambda_\phi(\phi^2 - \vt^2)^2 + \lambda_\psi\left[|\psi_1|^2 + 
  |\psi_2|^2
   - \wt^2+ g(\phi^2  - \vt^2)\right]^2
  \nonumber\\
  & & \quad {} + h |\psi_1|^2 |\psi_2|^2 - \mu\phi(|\psi_1|^2 - |\psi_2|^2)\ .
  \label{pot2}
\end{eqnarray}
In the vacuum the real scalar $\phi$ takes the vev $\pm v$
and there may exist domain walls separating these two values.  When
$\langle\phi\rangle=+v$, the vacuum has $|\psi_1| = v$ and $\psi_2 = 0$,
while when $\langle\phi\rangle=-v$ the vacuum has $|\psi_2| = v$ and  
$\psi_1 = 0$.  The values of $v$ and $w$ are as in the last section.

In this model, therefore, the unbroken symmetry group in the true
vacuum is $\u1$,
and the vacuum manifold is $\M = [\u1 \times\u1 \times \Z_2]/\u1
= S^1 \times \Z_2$, admitting walls and strings.  
When $\langle\phi\rangle=+v$, the complex field
$\psi_1$ can form cosmic strings with winding number $n$, around
which $\theta_1$ will change by $2\pi n$.  Such a string ends if
it intersects a D-wall, since $\langle\psi_1\rangle=0$ on the other 
side.  Analogous statements hold for the $\psi_2$ field when
$\langle\phi\rangle=-v$.

\begin{figure}
  \centerline{\epsfbox{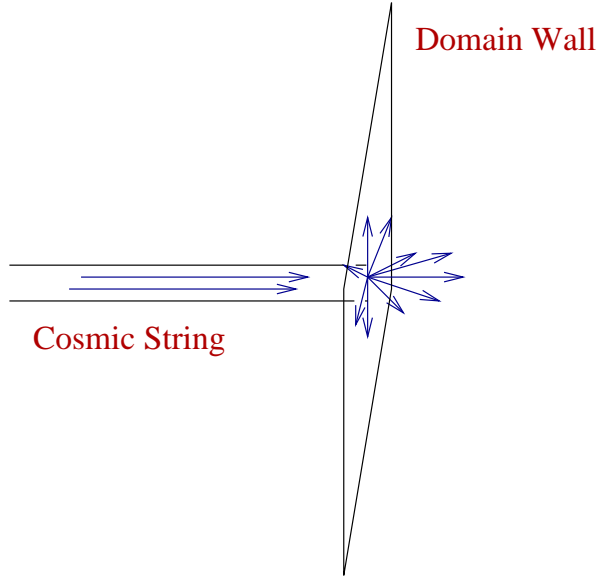}}
  \vspace{0.3in}
  \caption{\sf A cosmic string ends on a Dirichlet domain wall in 3 
  spatial dimensions. The arrows denote the magnetic flux.}
  \label{stwall}
\end{figure}

In the core of a string the
corresponding $\u1$ symmetry is restored.  In the gauge case,
therefore, the gauge bosons associated with, for example, ${\u1}_1$ 
are massless both in the core of a $\psi_1$-string on the
$\langle\phi\rangle = v$ side of the D-wall, and anywhere on the
$\langle\phi\rangle = -v$ side of the D-wall.  As usual, outside
the $\psi_1$-string the gauge field is pure gauge,
such that it cancels the gradient energy of the scalars by  
enforcing the vanishing of the covariant derivative
(\ref{covderiv}).  The gauge field is thus given by
$A_\mu^{(1)}  =  - \partial_\mu\theta_1$.
Consequently, there is magnetic flux through the string (which
we take to have winding number $n$), given by $\Phi^{(1)} = -n\pi$.
This flux flows through the string until it hits the wall; on the
other side of the wall the symmetry is unbroken everywhere, and the
magnetic field describes a monopole configuration emanating from
the point where the string intersects the wall.
We sketch such a configuration in Fig.~\ref{stwall}. 

Configurations of the this type, with strings ending on walls, have
recently been discussed in the context of supersymmetric QCD \cite{qcd}.
There, the string consists of non-Abelian flux, and the wall
separates different chiral vacua, with shifted values of the QCD
$\theta$-parameter.  The intersections of strings and domain walls
can be thought of as quarks.  
The structures of these QCD configurations
and the scalar field models discussed here are obviously very similar,
and the relationship between them deserves further investigation.
(One difference is that the flux in the strings considered in \cite{qcd}
does not propagate freely on the other side of the wall, as the symmetry
is still broken there; rather, it is confined to the wall itself.
It should not be difficult to extend models of the type considered in
this paper to include such situations.)  They have also been shown to
exist in certain grand unified theories \cite{dlv}.  Here, conventional
GUT monopoles can become bound to domain walls, and the monopoles will
become connected by strings if the color gauge group $\su3$ is
broken to $\Z_3$.

\section{Strings and strings}
\label{strings}

A number of theories in which a cosmic string can end on another
string can be found in the literature.  Generally speaking,
three-string vertices can arise whenever the strings are associated
with elements $a$, $b$, and $c$ of $\p1(\M)$, such that $abc= 1$.
Configurations of this type play an important role in models where
$\p1(\M)$ is non-Abelian, and may have interesting
cosmological consequences (see for example \cite{nonabelian}).

Nevertheless, it is interesting to consider theories of strings
ending on strings which more closely resemble those of the last
two sections; that is, with two types of strings, one of which may
be thought of as fundamental and the other as Dirichlet
(characterized by the fact that fundamental strings can end on
Dirichlet strings but not vice-versa).  As we shall see, the
construction of such models closely parallels that of the
theories with Dirichlet walls.

We once again consider three fields $\phi$, $\psi_1$, and $\psi_2$,
now with all three being complex scalars, for a total of six real degrees
of freedom.  We impose two $\u1$ symmetries, under which the fields have
the following charges:
\begin{equation}
  \matrix{ & & \underline{{\u1}_1} & \underline{{\u1}_2} \cr \cr
  \phi     & &   2    &    0    \cr
  \psi_1   & &    1    &    1    \cr
  \psi_2   & &    1    &   -1    \cr}
  \label{charges}
\end{equation}
We also include a $\Z_2$ with action
\begin{equation} 
  \psi_1 \leftrightarrow \psi_2\ .
\end{equation}
A general potential may be written in a form reminiscent of (but not
identical to) our previous examples:
\begin{eqnarray}
  V(\phi, \psi_1,\psi_2) & = & \lambda_\phi\left(|\phi|^2 - \vt^2\right)^2 +
  \lambda_\psi\left[|\psi_1|^2 + |\psi_2|^2 - \wt^2
  +g(|\phi|^2 - \vt^2)\right]^2  
  \nonumber\\
  & & \quad +h\left(|\psi_1|^2 - |\psi_2|^2\right)^2
  -\mu \left(\psi_1^*\psi_2^*\phi + \psi_1\psi_2\phi^*\right)\ .
  \label{sspot}
\end{eqnarray}

We describe the complex scalars in terms of their moduli and
phases as $\phi = \rho_\phi e^{i\theta_\phi}$,
$\psi_1 = \rho_1 e^{i\theta_1}$, $\psi_2 = \rho_2 e^{i\theta_2}$.
As before, for $\mu=0$ the potential is the sum of three
non-negative terms, which can be simultaneously set to zero by
setting $\langle\rho_\phi\rangle = \vt$ and 
$\langle\rho_1\rangle = \langle\rho_2\rangle = \wt$.
Turning on a small positive $\mu$ introduces a constraint on
the phases: $\langle\theta_\phi\rangle = \langle\theta_1\rangle 
+ \langle\theta_2\rangle$.
The unbroken subgroup is $\Z_2$, which in general is a combination
of the original $\Z_2$ and a ${\u1}_2$ transformation.  The vacuum
manifold is therefore a torus, $\M = [\u1 \times\u1 \times \Z_2]/\Z_2
= S^1 \times S^1$, which may be parameterized by the angles $\theta_1$
and $\theta_2$, determining the remaining angle $\theta_\phi$.

Strings are characterized by elements of $\p1(S^1\times S^1) =
\Z\times\Z$.  We can take the two generators $(1,0)$ and $(0,1)$ to be
given by paths from $\theta_i = 0$ to $\theta_i = 2\pi$, for $i=1,2$
respectively.  (In each case $\theta_\phi$ also ranges from $0$ to
$2\pi$.)  Due to the $\Z_2$ symmetry, strings corresponding to either
of the two generators have equal tensions.  A string
corresponding to $(1,-1)$, although it may be thought of as a
superposition of strings with charges $(1,0)$ and $(0,-1)$, can have a
lower energy than the two separately (since $\rho_1$ equals
$\rho_2$, and thus the term in the potential of the form
$h[|\psi_1|^2-|\psi_2|^2]$ vanishes) and therefore be stable
against decay; strings of this type are the fundamental
strings, while those with charges $(1,0)$ or $(0,1)$ are
the D-strings.

The gauge transformations in $\u1_1$ and $\u1_2$ can be written in terms
of functions $\omega_1(x^\mu)$ and $\omega_2(x^\mu)$ as
\begin{equation}
  \left(\matrix{\phi \cr \psi_1 \cr \psi_2\cr}\right)
  \rightarrow
  \left(\matrix{ e^{-2i\omega_1}\phi \cr 
  e^{-i(\omega_1+\omega_2)}\psi_1 \cr 
  e^{-i(\omega_1-\omega_2)}\psi_2\cr}\right)\ .
  \label{u1u1tx}
\end{equation}
In the core of the $(1,-1)$ string, we have $\langle\phi\rangle= v$,
$\langle\psi_1\rangle = \langle\psi_2\rangle = 0$, and the
$\u1_2$ symmetry (transformations with $\omega_1=0$) is restored.
In a $(1,0)$ string there is an unbroken $\u1$ parameterized by
$\omega_1 = \omega_2$, and in a $(0,1)$ string there is an unbroken 
$\u1$ parameterized by $\omega_1 = -\omega_2$.

Outside the string cores, the gauge fields are pure gauge 
such that they cancel the gradient energy in the scalars.  This 
means they enforce the vanishing of the covariant derivatives
\begin{eqnarray}
 D_\mu\phi & = & \partial_\mu\phi +2i A_\mu^{(1)}\phi\ ,
 \nonumber\\
 D_\mu\psi_1 & = & \partial_\mu\psi_1 +i A_\mu^{(1)}\psi_1
  + i A_\mu^{(2)}\psi_1\ ,
 \\
 D_\mu\psi_2 & = & \partial_\mu\psi_2 +i A_\mu^{(1)}\psi_2
  - i A_\mu^{(2)}\psi_2\ .
  \nonumber
\end{eqnarray}
The gauge fields are thus given by
\begin{eqnarray}
  A_\mu^{(1)} & = & {1\over{2}} \partial_\mu(\theta_1 + \theta_2)
  = {1\over{2}} \partial_\mu\theta_\phi\ ,
  \nonumber\\
  A_\mu^{(2)} & = & {1\over{2}} \partial_\mu(\theta_1 - \theta_2)
  \ .
\end{eqnarray}
Consequently, the magnetic fluxes through a string with winding
number $(n,m)$ are
\begin{eqnarray}
  \Phi^{(1)}(n,m) & = & (n+m)\pi\ ,
  \nonumber\\
  \Phi^{(2)}(n,m) & = & (n-m)\pi\ .
\end{eqnarray}

\begin{figure}
  \centerline{\epsfbox{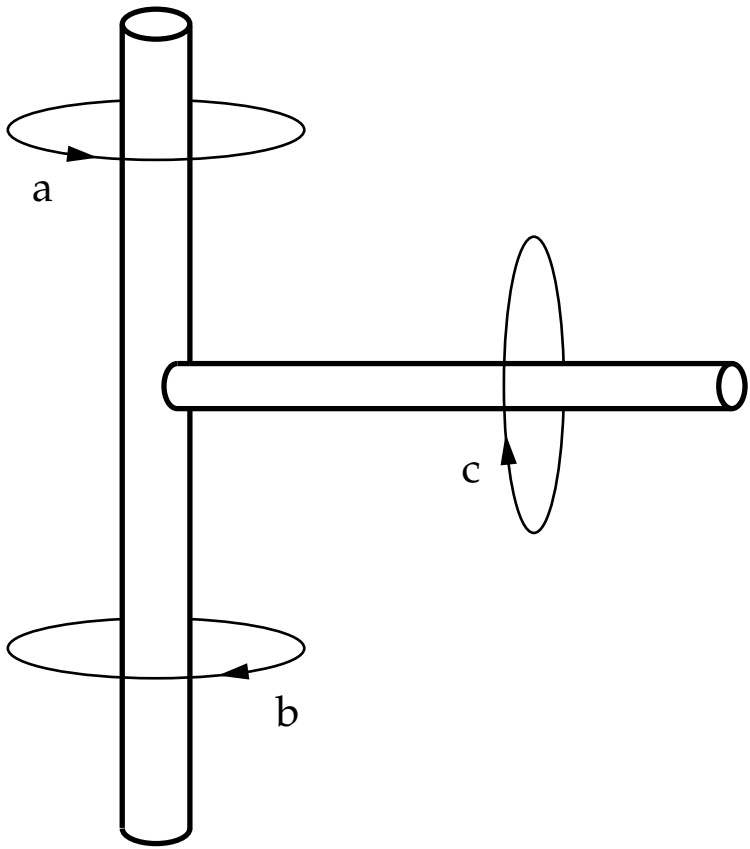}}
  \vspace{0.3in}
  \caption{\sf A fundamental cosmic string ends on a Dirichlet string.  
  The homotopy classes of the indicated loops in 
  $\p1(S^1\times S^1) = \Z\times\Z$ are $[a] = (1,0)$, $[b] = (0,-1)$, and 
  $[c] = (1,-1)$.}
\label{ssfig}
\end{figure}

In Fig.~\ref{ssfig} we show a configuration with a fundamental string
ending on a D-string.  Around the latter, the phase of $\phi$ changes
by $2\pi$; above the intersection point it is accompanied by the phase
of $\psi_1$ ($\theta_1 = \theta_\phi$, $\theta_2 = 0$), while below
the intersection point it is accompanied by the phase of $\psi_2$
($\theta_2 = \theta_\phi$, $\theta_1 = 0$).  The fundamental string is
topologically equivalent to the superposition of the two segments of
the D-string (as it must be, since the loops $a$ and $b$ together are
topologically equivalent to loop $c$); around it the phases $\theta_1$
and $\theta_2$ change by $2\pi$, while $\theta_\phi$ is
constant.
It is easy to verify that the configuration shown conserves flux:
through $a$ we have $\Phi^{(1)} = \pi$, $\Phi^{(2)} = \pi$; 
through $b$ we have $\Phi^{(1)} = \pi$, $\Phi^{(2)} = -\pi$; and 
through $c$ we have $\Phi^{(1)} = 0$, $\Phi^{(2)} = 2\pi$.

\section{Conclusions}
\label{concl}

We have described a class of topological defects in
classical field theories in (3+1) dimensions, consisting of Dirichlet
defects on which fundamental defects of lower dimension can
terminate.  While the search for models supporting
these configurations is inspired by the appearance of D-branes in
string theory, there are important differences between the two
sets of objects.  In all of the theories we consider, the
basic degrees of freedom are scalar fields and gauge fields,
out of which all of the higher-dimensional objects are constructed.
Our theories do not include the effects of gravity, and are not
supersymmetric (although there are no obstacles to the appropriate
generalizations).  Furthermore, the specific dependence of D-brane
energy on the string coupling constant is not a feature of our models,
and the Ramond-Ramond gauge fields to which D-branes couple are
absent.  Nevertheless, it may be interesting to compare the
dynamical behavior of Dirichlet defects to that of D-branes in
string theory, and search for models in which the similarities 
between the two systems are even stronger.

One obvious direction in which to generalize the models considered
here is to consider $q$-dimensional defects ending on
$p$-dimensional D-defects in $d$ spatial dimensions.  (There are
a variety of such objects in string theory and M-theory, with
configurations governed by charge conservation \cite{strom}.)
A number of
interesting issues arise in this case, especially for gauge
symmetries.  In three spatial dimensions, domain walls, strings and
monopoles all have finite energy (per unit volume) if they arise
from the spontaneous breakdown of gauge symmetries, while 
strings and monopoles in theories with global symmetries have
divergent energies.  More generally, gauge defects of codimension 
greater than three have divergent energies, as have global
defects of codimension greater than one \cite{coleman}.  Although
the divergence of the energy is an important feature of such
configurations, it does not render them unphysical; certainly there
is no obstacle in principle, for example, to the existence of
global strings in (3+1) dimensions.  More importantly, to make
topological defects of dimension $q$ in $d$ spatial dimensions
requires that $\p{d-q-1}(\M)$ be nontrivial, for example by
breaking $\so{d-p}$ to $\so{d-p-1}$ (for which $\M = S^{d-p-1}$).
In such a model, the unbroken symmetry group $\so{d-p-1}$ is
non-Abelian for $p\leq d-4$; we then expect the low-energy gauge theory
to be strongly coupled, and the resulting defects to be confined.

Back in (3+1) dimensions, there are a number of issues remaining
to be addressed.  As mentioned, the models we have 
constructed are purely bosonic, and it would be interesting to
consider supersymmetric versions 
(as has been done for non-hybrid defects \cite{susydefects}), 
as well as to determine whether
Dirichlet defects could arise in realistic particle physics
models.  Finally, as with any species of topological defect,
it is also natural to ask what the cosmological consequences of
the formation of these objects in the early universe might be.

\section*{Acknowledgments}

We would like to thank Andrew Chamblin, Gary Gibbons, Aki Hashimoto,
Miguel Ortiz, Joe Polchinski, John Preskill, Wati Taylor and Tanmay
Vachaspati for helpful conversations.  The work of S.M.C. was
supported in part by the National Science Foundation under grant
PHY/94-07195 and the work of M.T. was supported by the Department of
Energy (D.O.E.), the National Science Foundation (N.S.F.) and by funds
provided by Case Western Reserve University.

\end{document}